\documentclass[twocolumn,twoside,preprintnumbers,amsmath,amssymb,pacs]{revtex4}
\usepackage{epsfig}
\usepackage{graphicx}
\usepackage{dcolumn}
\usepackage{bm}

\usepackage{fancyhdr}

\usepackage{pslatex}

\newcommand{\rr}{\mbox{\boldmath $r$}}

\def\gappeq{\mathrel{\rlap {\raise.5ex\hbox{$>$}}
{\lower.5ex\hbox{$\sim$}}}}
\def\beq{\begin{equation}}
\def\eeq{\end{equation}}
\def\bea{\begin{eqnarray}}
\def\eea{\end{eqnarray}}
\def\bq{\begin{quote}}
\def\eq{\end{quote}}
\def\lfrestriction#1{\lower.25ex\hbox{\Big|}_{#1}}

\begin{document}

\title{{\Large  Neutrino structure functions in the QCD dipole picture}}

\author{M.B. Gay Ducati, M.M. Machado}

\affiliation{High Energy Physics Phenomenology Group, GFPAE,  IF-UFRGS \\
Caixa Postal 15051, CEP 91501-970, Porto Alegre, RS, Brazil}

\author{M.V.T.  Machado}

\affiliation{Centro de Ci\^encias Exatas e Tecnol\'ogicas, Universidade Federal do Pampa \\
Campus de Bag\'e, Rua Carlos Barbosa. CEP 96400-970. Bag\'e, RS, Brazil}


\received{on 30 October, 2006}

\begin{abstract}
In this contribution we present an exploratory QCD analysis of the neutrino structure functions in charged current DIS using the color dipole formalism. The corresponding dipole cross sections are taken from recent phenomenological and theoretical studies in deep inelastic inclusive production, including nuclear shadowing corrections. The theoretical predictions are compared to the available experimental results in the small-$x$ region. \\


\end{abstract}

\maketitle

\thispagestyle{fancy}
\setcounter{page}{0}

\section{Introduction}

The interaction of  high energy neutrinos on hadron targets are an outstanding probe to test Quantum Chromodynamics (QCD) and understanding the parton properties of hadron structure. The several combinations of neutrino and anti-neutrino scattering data can be used to determine the structure functions, which constrain the valence, sea and gluon parton distributions in the nucleons/nuclei. The comparison between neutrino and charged-lepton experimental data can be also used to investigate the universality of the parton distributions. The differential cross section for the neutrino-nucleon charged current process $\nu_l\,(\bar{\nu}_l)+N \rightarrow l^-\,(l^+)+X$, in terms of the Lorentz invariant structure functions $F_2^{\nu N}$, $2xF_1^{\nu N}$ and $xF_3^{\nu N}$ are \cite{Leader_Predazzi},
\begin{eqnarray}
\frac{d\sigma^{\nu ,\,\bar{\nu}}}{dx \,dy} & = & \frac{G_F^2 \,m_N \,E_{\nu}}{\pi }\left[ \left( 1-y-\frac{m_Nxy}{2E_{\nu}} \right) F_2(x,Q^2) \right. \nonumber \\
& + &\left. \frac{y^2}{2} \,2xF_1(x,Q^2)\pm y\left( 1-\frac{y}{2} \right) xF_3(x,Q^2) \right]\,,
\end{eqnarray}
where $G_F$ is the weak Fermi coupling constant, $m_N$ is the nucleon mass, $E_{\nu}$ is the incident neutrino energy, $Q^2$ is the square of the four-momentum transfer to the nucleon. The variable $y=E_{had}/E_{\nu}$ is the fractional energy transferred to the hadronic vertex with $E_{had}$ being the measured hadronic energy, and $x=Q^2/2m_NE_{\nu}y$ is the Bjorken scaling variable (fractional momentum carried by the struck quark).

Similarly to the charged-lepton DIS, the deep inelastic neutrino scattering is also used to investigate the structure of nucleons and nuclei. In the leading order quark-parton model (the QCD collinear approach), the structure function $F_2$ is the singlet distribution, $F_2^{\nu N}\propto xq^S=x\sum(q+\bar{q})$, the sum of momentum densities of all interacting quarks constituents, and $xF_3$ is the non-singlet distribution, $xF_3^{\nu N}\propto xq^{NS}=x\sum(q-\bar{q})=xu_V+xd_V$, the valence quark momentum density. These relations are further modified by higher-order QCD corrections. The main theory uncertainties are the role played by nuclear shadowing in contrast with lepton-charged DIS and a correct understanding of the low $Q^2$ limit. The first uncertainty can be better addressed with the future precise data from MINER$\nu$A \cite{minerva} and neutrino-factory \cite{nufactory}. However, nuclear effects are taken into account by using the nuclear ratios $R=F_2^A/AF_2^p$ extracted from lepton-nucleus DIS, which could be different for the neutrino-nucleus case. The low-$Q^2$ region can not be addressed within the pQCD quark-parton model as a hard momentum scale $Q_0^2\geq 1-2$ GeV$^2$ is required in order to perform perturbative expansion. In what follows, the neutrino structure function $F_2^{\nu N}$ is investigated within the color dipole picture at small-$x$ region. The present calculations are discussed in detail in Ref. \cite{GMM}. We employ recent phenomenological parton saturation models, which are successful in describing inclusive deep inelastic data. Nuclear effects are taken into account through Glauber-Gribov formalism and the results are compared to accelerator data. Afterwards, the structure function $xF_3$ and the  quantity $\Delta x F_3=xF_3^{\nu}-xF_3^{\bar \nu}$ are also investigated. The latter quantity provides a determination of the strange-sea parton distribution through charm production in charged-current neutrino DIS. Finally, we present a brief summary of the main conclusions.

\section{Neutrino Structure Function $F_2^{\nu N}$}

Deep inelastic neutrino scattering can proceed via  $W^{\pm}$ exchanges, which corresponds to charged current (CC) interactions. We assume an isoscalar target, $N=(p+n)/2$ and focus on the high energy regime, which one translates into small-$x$ kinematical region. At this domain a quite successful framework to describe QCD interactions is provided by the color dipole formalism \cite{DIPOLEPIC}, which allows an all twist computation  (in contrast with the usual leading twist approximation) of the structure functions. The physical picture of the interaction is the deep inelastic scattering  at low $x$  viewed as the result of the interaction of a $q \bar{q}$ color dipole, in which the gauge boson fluctuates into, with the nucleon target. The interaction is modeled via the dipole-target cross section, whereas the boson fluctuation in a color dipole is given by the corresponding wave function. The CC  DIS structure functions for neutrino-nucleon scattering in the dipole picture \cite{GMM} are related to the cross section for scattering of transversely and longitudinally polarized $W^{\pm}$ bosons. That is,
\begin{eqnarray}
F_{T,L}^{\mathrm{CC}}(x,Q^2) = \frac{Q^2}{4\,\pi^2}\int d^2 \rr \int_0^1 dz
| \psi^{W^{\pm}}_{T,L}(z,\rr,Q^2)|^2\sigma_{dip}(x,\rr) ,
\label{FSDIP}
\end{eqnarray}
where  $\rr$ denotes the transverse size of the color dipole, $z$ the
longitudinal momentum fraction carried by a quark and  $\psi^{W}_{T,L}$ are  the light-cone wavefunctions for (virtual) charged gauge bosons with transverse or longitudinal polarizations. The small-$x$ neutrino structure function $F_2^{\nu N}$ is computed from expressions above taking $F_2=F_T+F_L$. Explicit expressions for the wave functions squared  can be found in Ref. \cite{GMM}.  The color dipoles  contributing  to Cabibbo favored transitions are  $ u \bar d \, (d \bar u)$,  $ c \bar s \,(s \bar c)$ for CC interactions. The dipole hadron cross section $\sigma_{dip}$  contains all
   information about the target and the strong interaction physics.

We consider an analytical expression for the dipole cross section, which presents scaling behavior. Namely, one has $\sigma_{dip} \propto (\rr^2Q^2_{\mathrm{sat}})^\gamma$ for dipole sizes $\rr^2 \approx 1/Q^2_{\mathrm{sat}}$ and where $\gamma$ is the effective anomalous dimension. The so-called saturation scale $Q_{\mathrm{sat}} \propto x^{\lambda/2}$ defines the onset of the parton saturation effects. In what follows one takes the phenomenological parameterization of Itakura-Iancu-Munier (IIM) model \cite{IIM}. It is able to describe experimental data on inclusive and diffractive deep inelastic $ep$ scattering at small-$x$. The IIM  model smoothly interpolates between the  limiting behaviors analytically under control: the solution of the BFKL equation
for small dipole sizes, $\rr\ll 1/Q_{\mathrm{sat}}(x)$, and the Levin-Tuchin prediction for larger ones, $\rr\gg 1/Q_{\mathrm{sat}}(x)$. A fit to the structure function $F_2(x,Q^2)$ was performed in the kinematical range of interest.  The IIM dipole cross section is parameterized as follows,
\begin{eqnarray}
\sigma_{dip}^{\mathrm{IIM}}\,(x,\rr) =\sigma_0\,\left\{
\begin{array}{ll}
{\mathcal N}_0 \left(\frac{\bar{\tau}^2}{4}\right)^{\gamma_{\mathrm{eff}}\,(x,\,r)}\,, & \mbox{for $\bar{\tau} \le 2$}\,, \nonumber \\
 1 - \exp \left[ -a\,\ln^2\,(b\,\bar{\tau}) \right]\,,  & \mbox{for $\bar{\tau}  > 2$}\,,
\end{array} \right.
\label{CGCfit}
\end{eqnarray}
where $\bar{\tau}=\rr Q_{\mathrm{sat}}(x)$ and the expression for $\rr Q_{\mathrm{sat}}(x)  > 2$  (saturation region)   is an adequate functional
form, compatible with approximate or asymptotic solutions of the high energy evolution equations.  The coefficients $a$ and $b$ are determined from the continuity conditions of the dipole cross section  at $\rr Q_{\mathrm{sat}}=2$. The coefficients $\gamma_{\mathrm{sat}}= 0.63$ (the BFKL anomalous dimension at the saturation border) and $\kappa= 9.9$  are fixed from their LO BFKL values. The IIM parameterization presents scaling violation since the effective anomalous dimension depends also on the rapidity $Y=\ln(1/x)$ for small size dipoles, $\gamma\,(x,\rr)=\gamma_{\mathrm{sat}} + \frac{\ln (2/\rr Q_{\mathrm{sat}})}{\kappa \,\lambda \,Y}$. Using the dimensional-cutting rules, we supplement
 the dipole cross section with a threshold factor
$(1-x)^{n_{\mathrm{thres}}}$, taking $n_{\mathrm{thres}}=5$.

The extension of the approach to consider nuclei targets we take the Glauber-Gribov picture \cite{ARMESTO}, without any new parameter. In this
approach, the nuclear version is obtained replacing the
dipole-nucleon cross section  by the
nuclear one,
\begin{eqnarray}
\sigma_{dip}^{\mathrm{nucleus}} = 2\,\int d^2b \,
\left\{\, 1- \exp \left[-\frac{1}{2}\,T_A(b)\,\sigma_{dip}^{\mathrm{nucleon}} (\bar{x}, \,\rr^2)  \right] \, \right\}\,,
\label{sigmanuc}
\end{eqnarray}
where $b$ is the impact parameter of the center of the dipole
relative to the center of the nucleus and the integrand gives the
total dipole-nucleus cross section for a  fixed impact parameter.
The nuclear profile function is labeled by $T_A(b)$, which will
be obtained from a 3-parameter Fermi distribution for the nuclear
density  \cite{devries}.


\begin{figure}[t]
\centerline{\includegraphics[scale=0.5]{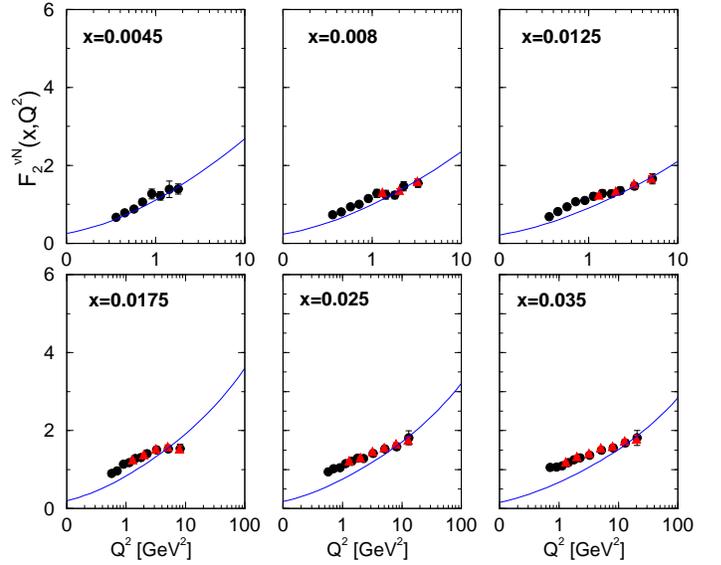}}
\caption{The structure function $F_2^{\nu N}$ as a function of boson virtuality. }
\label{fig:1}
\end{figure}

Let us compare the color dipole prediction against the $F_2^{\nu N}$ structure function. This is shown in Fig. \ref{fig:1}. We use the experimental datasets of the CCFR Collaboration \cite{CCFR,Fleming}, where filled circles correspond to points in Ref. \cite{Fleming} and triangles up correspond to points in Ref. \cite{CCFR}. The solid curve is obtained using scaling property and nuclear shadowing from Glauber-Gribov formalism is also included (estimated to be of order 20\% at small-$x$). The data description is reasonable up to $x\simeq 0.0175$, just in the border of the expected validity region of the color dipole approach. For completeness, we show the larger $x$ data points. The valence content has not been included and it could improve the description in that region. It has been verified in Ref. \cite{GMM} that the improvement is sizable at low-$Q^2$. On the other hand, the large $Q^2$ region tends to be  overestimated for $x>0.0175$. Thus, a more detailed study of that region is deserved. It is worth to mention the robustness of the color dipole formalism as the theoretical curves were obtained without any tuning of the original model parameters obtained in $ep$ HERA data. In addition, we have shown in Ref. \cite{GMM} that the small-$x$ data on $F_2$ can be used to verify that geometric scaling  property is exhibited by experimental results.

\section{The functions $xF_3^{\nu N}$ and the quantity $\Delta xF_3^{\nu N}$ }

Lets now compute the structure functions $xF_3^{\nu N}$ within the color dipole formalism. We concentrate on the the
interaction of the $c\bar s$ color dipole of size $\rr$
 with the target  hadron which is described by the beam-
 and flavor-independent color dipole cross section
$\sigma_{dip}$. In the infinite momentum frame, this is equivalent to the $W^{\pm}$-gluon fusion process, $W^{\pm}+g\rightarrow  c\bar{s} \,(\bar{c}s)$.
The analysis for charged current DIS has been addressed in Refs. \cite{ZOLLER1,ZOLLER2}, where the left-right asymmetry  of diffractive
interactions of electroweak bosons of different helicity is discussed. There, the relevant light-cone wavefunctions have been evaluated. The contribution of excitation of open
charm/strangeness to the hadron absorption cross section for
left-handed ($L$)  and right-handed ($R$) $W$-boson of virtuality $Q^2$,
is given by \cite{GMM},
\begin{eqnarray}
\sigma_{L,\,R}\,(x,Q^{2})
=\int d^{2}{\rr} \int_0^1 dz \sum_{\lambda_1,\lambda_2}
|\Psi_{L,\,R}^{\lambda_1,\lambda_2}(z,\rr,Q^2)|^{2}
\,\sigma_{dip}\,(x,\rr )\,,
\label{eq:FACTOR}
\end{eqnarray}
where $\Psi_{L,\,R}^{\lambda_1,\lambda_2}(z,\rr,Q^2)$
 is the light-cone wavefunction of
the $c\bar{s}$ state with the $c$ quark
carrying fraction $z$ of the $W^+$ light-cone momentum and
$\bar s$ with momentum fraction $1-z$. The $c$- and $\bar s$-quark
helicities are  $\lambda_1=\pm 1/2$ and  $\lambda_2=\pm 1/2$, respectively.
The diagonal elements of  density matrix are computed in Ref. \cite{ZOLLER1,ZOLLER2}.

\begin{figure}[t]
\centerline{\includegraphics[scale=0.46]{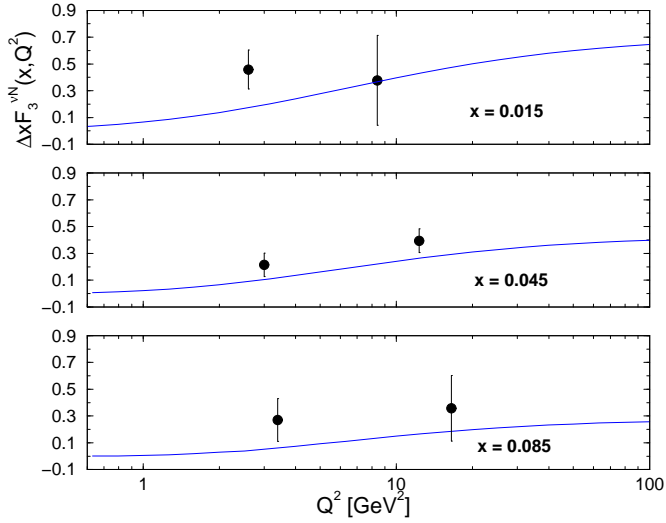}}
\caption{The structure function $\Delta xF_3^{\nu N}=xF_3^{\nu N}-xF_3^{\bar{\nu} N}$ as a function of boson virtuality.}
\label{fig:2}
\end{figure}

The structure function of deep inelastic neutrino-nucleon  $xF_3$ can be defined in terms of $\sigma_{R}$ and $\sigma_{L}$ of Eq.~(\ref{eq:FACTOR}) in the following usual way,
\begin{eqnarray}
xF_3^{\nu N}\,(x,Q^2)=\frac{Q^2}{ 4\pi^2}
\left[\sigma_{L}(x,Q^{2})-\sigma_{R}(x,Q^{2})\right].
\label{eq:F3}
\end{eqnarray}
where the expression can be interpreted in terms
of parton densities  as being the sea-quark component
of $xF_3$.  It corresponds to the excitation of the $c\bar s$ state in the process $W^+g\rightarrow c\bar{s}$, with $xF_3$
differing from zero due to the strong left-right asymmetry of the
light-cone $|c\bar s\rangle$ Fock state. For values of Bjorken variable not so small, $xF_3$ contains important valence quark contribution. The valence term, $xq_{val}$, is the same for both  $\nu N$ and
$\bar\nu N$ structure functions of an  iso-scalar nucleon.
The sea-quark ($xq_{sea}$) term in the $xF^{\nu N}_3$ has opposite
sign for $xF^{\bar\nu N}_3$, leading to $xF^{\,\nu (\bar{\nu}) N}_3=xq_{val}\pm xq_{sea}$. The neutrino-antineutrino difference $xF_3^{\nu}-xF_3^{\bar{\nu}}$ provides a determination of the sea (strange) density. In the parton model, one has $xF^{\nu N}_3=xq_{val} -2x\bar{c}(x)+2xs(x)$ and $xF^{\bar{\nu} N}_3=xq_{val}+ 2xc(x)-2x\bar{s}(x)$. Therefore, the neutrino-antineutrino difference effectively measures the strange density, since the charm contribution is small in the kinematical region measured by current experiments. Assuming $s(x)=\bar{s}(x)$ and $c(x)=\bar{c}(x)$ one obtains, $\Delta x F_3=xF_3^{\nu N}-xF_3^{\bar \nu N}=2xq_{sea}= 4x\left[s(x)-c(x)\right]$. In Fig. \ref{fig:2} the quantity  $\Delta x F_3$ as a function of $Q^2$ at fixed $x$ is shown in comparison with the CCFR result obtained from $\nu_{\mu}Fe$ and $\bar\nu_{\mu}Fe$ differential cross section \cite{CCFR3}.  The theoretical curve is obtained from the corresponding equation using the IIM dipole cross section and Glauber-Gribov shadowing corrections.

As a summary, an analysis of small-$x$ neutrino-nucleus DIS is performed within the color dipole formalism. The structure functions $F_2^{\nu N}$,  $xF_3^{\nu N}$ and the quantity $\Delta xF_3^{\nu N}$ are calculated and compared with the experimental data from CCFR and NuTeV by employing  a recent parameterization for the dipole cross section which successfully describe small-$x$ inclusive and diffractive $ep$ DIS data. Nuclear shadowing is taking into account through Glauber-Gribov formalism. The structure function $F_2$ is in agreement with an implementation from saturation models at the region $x \leq 0.02$. This is in agreement with the regime of validity of the color dipole approach. The sea content, described by the quantity $\Delta xF_3^{\nu N}$, is well described. Although the results presented here are compelling, further investigations are requested.



\begin{thebibliography}{}

\bibitem{Leader_Predazzi} E. Leader and E. Predazzi, An Introduction to Gauge Theories and the New Physics (Cambridge Univ. Press., Cambridge, 1982).

\bibitem{minerva} D. Drakoulakos et al. [Minerva Collaboration], FERMILAB-PROPOSAL-0938; arXiv:hep-ex/0405002.

\bibitem{nufactory} S. Kumano, arXiv:hep-ph/0310166.

\bibitem{DIPOLEPIC}
A. H. Mueller,  Nucl. Phys. {\bf B335} (1990) 115;
N.N. Nikolaev and B.G. Zakharov,  Z. Phys. {\bf C49}
(1991) 607.

\bibitem{ARMESTO}
N. Armesto, Eur. Phys. J. {\bf C26} (2002) 35.


\bibitem{GMM} M.B. Gay Ducati, M.M. Machado, M.V.T. Machado, Phys. Lett. B {\bf 644}, 340 (2007).


\bibitem{IIM} E.~Iancu, K. Itakura and S. Munier, Phys.\ Lett.\ B {\bf 590}, 199 (2004).


\bibitem{devries}
C. W. De Jager, H. De Vries, C. De Vries, Atom. Data Nucl. Data Tabl. {\bf 14}, 479 (1974).


\bibitem{CCFR}
W.G. Seligman et al. [CCFR Coll.], Phys. Rev. Lett. {\bf 79} (1997) 1213.

\bibitem{Fleming}
B.T. Fleming et al. [CCFR Coll.],  Phys. Rev. Lett. {\bf 86} (2001) 5430.

\bibitem{ZOLLER1} R. Fiore and V.R. Zoller, JETP Lett. {\bf 82} (2005) 385.

\bibitem{ZOLLER2} R. Fiore and V.R. Zoller, Phys. Lett. {\bf B632} (2006) 87.

\bibitem{CCFR3}
U.K. Yang et al.,  Phys. Rev. Lett. {\bf 86} (2001) 2742.

\end{thebibliography}
\end{document}